\documentclass[aps,prb,twocolumn,showpacs,groupedaddress]{revtex4-1}
\usepackage{amssymb}

\usepackage{graphicx}

\begin{document}

\title[]{Magnetic and multiferroic phases of single-crystalline Mn$_{0.85}$Co$_{0.15}$WO$_4$}

\author{R. P. Chaudhury$^1$, F. Ye$^2$, J. A. Fernandez-Baca$^{2,3}$, Y.-Q. Wang$^1$, Y. Y. Sun$^1$, B. Lorenz$^1$,  H. A. Mook$^2$, and C. W. Chu$^{1,4}$}

\affiliation{$^1$ TCSUH and Department of Physics, University of Houston, Houston, Texas 77204-5002, USA}

\affiliation{$^2$ Neutron Scattering Science Division, Oak Ridge National Laboratory, Oak Ridge, TN 37831-6393, USA}

\affiliation{$^3$ Department of Physics and Astronomy, The University of Tennessee, Knoxville, TN 37996-1200, USA}

\affiliation{$^4$ Lawrence Berkeley National Laboratory, 1 Cyclotron Road, Berkeley, California 94720, USA}

\begin{abstract}
The magnetic and multiferroic phase diagram of Mn$_{0.85}$Co$_{0.15}$WO$_4$ single crystals is investigated by means of magnetic, heat capacity, dielectric, polarization, and neutron scattering experiments. Three magnetic phase transitions are detected through distinct anomalies in all physical quantities. The ferroelectric polarization is observed only along the b-axis below 10 K but not along the a-axis as recently suggested. The magnetic phases studied by neutron scattering are very complex. Up to four different magnetic structures, partially coexisting at certain temperature ranges, have been identified. Upon decreasing temperature two commensurate phases (AF4, AF1) are followed by an incommensurate phase (AF5) and a second incommensurate phase (AF2) is detected as a minor phase. The ferroelectric polarization is possibly associated with both (AF2 and AF5) phases.
\end{abstract}


\maketitle

\section{Introduction}
The renewed and increasing interest in multiferroic magnetoelectric materials in recent years has led to the discovery of new magnetoelectric compounds, novel physical properties, and to a better understanding of the fundamental physical processes and interactions leading to the complex behavior of many multiferroics. \cite{fiebig:05,spaldin:05,tokura:07} While different physical interactions are now known to result in the coexistence of ferroelectricity with magnetic or charge orders one of the more frequently observed and discussed mechanisms is based on an inversion symmetry breaking magnetic order that results in ionic displacements with a macroscopic electric polarization.\cite{mostovoy:06,mochizuki:10} For example, helical magnetic orders violating the inversion symmetry have been observed in the multiferroic state of TbMnO$_3$,\cite{kenzelmann:05} Ni$_3$V$_2$O$_8$,\cite{lawes:05} and MnWO$_4$.\cite{taniguchi:06} The magnetic system is usually highly frustrated due to geometric constraints and/or competing exchange interactions and the magnetic phases can be very complex including incommensurate collinear and non collinear structures as well as spin-frustrated commensurate phases. With sufficiently strong spin-lattice coupling any inversion symmetry breaking order (such as the transverse spin helix) may produce a polarized ferroelectric (FE) state. The high degree of frustration also explains the sensitivity of the multiferroic state with respect to external fields,\cite{taniguchi:06,higashiyama:04,hur:04,seki:08} pressure,\cite{delacruz:07,delacruz:08,chaudhury:07,chaudhury:08b} or chemical substitutions.\cite{chaudhury:08,chaudhury:09b,seki:07,kanetsuki:07}

MnWO$_4$ (mineral H\"{u}bnerite) passes through three magnetic phase transitions upon decreasing temperature (T). Below T$_N$=13.5 K a sinusoidal spin order with an incommensurate (IC) modulation, $\overrightarrow{q}_{3}=(-0.214,1/2,0.457)$, defines the AF3 phase. The Mn spins form an angle of about 34$^\circ$ with the a-axis lying in the a-c plane. The AF3 phase is followed by the AF2 phase at 12.6 K with a helical magnetic order. The spins tilt toward the b-axis to form a helix but the propagation vector remains unchanged, $\overrightarrow{q}_{2}$=$\overrightarrow{q}_{3}$. The AF2 phase is ferroelectric (multiferroic). At lower T $<$ 7.8 K the collinear AF1 phase with the characteristic $\uparrow\uparrow\downarrow\downarrow$ spin structure and the commensurate (CM) modulation vector $\overrightarrow{q}_{1}=(\pm1/4,1/2,1/2)$ becomes stable and represents the ground state. The magnetic, dielectric, and thermodynamic properties of MnWO$_4$ have been extensively investigated\cite{ehrenberg:97,taniguchi:06,arkenbout:06,taniguchi:08,chaudhury:08c,kundys:08} and neutron scattering experiments have revealed the details of the magnetic orders.\cite{lautenschlager:93,sagayama:08} The strong magneto-electric interaction in this compound results in novel phenomena such as the electric-field control of the chirality of magnetic domains\cite{finger:10} and the coupling of magnetic and ferroelectric domains.\cite{meier:09}

The effects of substitutions on the magnetic phases of MnWO$_4$ have been studied for the system Mn$_{1-x}$Fe$_x$WO$_4$.\cite{obermayer:73,klein:74,garciamatres:03,ding:00} The multiferroic and ferroelectric AF2 phase is completely suppressed by less than 5 \% Fe substitution but it can be restored in external magnetic fields  resulting in a complex field-temperature-composition phase diagram with competing and partially coexisting phases.\cite{chaudhury:08,ye:08,chaudhury:09b} At higher Fe substitution the sinusoidal AF3 phase transformed directly into the low-T AF1 phase. In contrast, substituting Co for Mn seems to stabilize the multiferroic AF2 phase and, at higher doping levels, the appearance of another helical magnetic structure (AF2') was reported in recent neutron scattering experiments on powder samples of Mn$_{1-x}$Co$_x$WO$_4$.\cite{song:09} It was further predicted that the AF2' phase gives rise to a large FE polarization primarily directed along the a-axis.

In order to investigate the multiferroic properties and the suggested spin- and polarization flop in detail the study of single crystals of the system Mn$_{1-x}$Co$_x$WO$_4$ is necessary. We have therefore grown large single crystals of Mn$_{0.85}$Co$_{0.15}$WO$_4$ in a floating zone optical furnace and investigated the magnetic and multiferroic phases employing magnetization, heat capacity, dielectric (polarization), and neutron scattering measurements.

\section{Experimental}
The typical size of the floating-zone grown crystals exceeds 5 mm in diameter and 10 mm in length. The crystals were oriented and cut according to the requirements of the different experiments. The chemical composition and uniformity of the crystals used for measurements have been verified by ICP-MS analysis of 10 different spots on the crystal's surface. The Co concentration was found close to the nominal value, x$_{Co}$=0.156$\pm$0.006. The small standard deviation shows the excellent uniformity of the crystal. The magnetic susceptibility and the heat capacity were measured in a superconducting quantum interference device (Quantum Design) and the Physical Property Measurement System (Quantum Design), respectively. The dielectric constant and the loss factor were estimated from the capacitance of a thin plate-like crystal employing the AH2500A capacitance bridge (Andeen Hagerling). Silver paint was applied to both sides of the platelet to form electrical contacts. the sample thickness was 0.5 mm and the contact area was 15 mm$^2$. The electric polarization was determined from the pyroelectric current measured by a K6517A electrometer (Keithley). The measured current was numerically integrated from high temperature (above T$_C$) to the lowest T to determine the temperature dependence of the ferroelectric polarization. The samples were cooled to the lowest temperature with an electrical bias field applied, kept at the lowest temperature with the contacts shortened for several minutes, and heated in zero bias field while recording the pyroelectric current. Special care had been taken to eliminate any artificial contribution to the current measurements in the ferroelectric state. The bias electric field (cooling) was 3 kV/cm, high enough to ensure the alignment of the FE domains. This was verified by measuring the FE polarization after cooling in the reduced electric field of 2 kV/cm and verifying the same magnitude of the FE polarization. Some experiments had been conducted in cooling in a negative (sign-reversed) electric field and the obtained polarization in the FE state was of equal magnitude but of opposite sign, as expected in a FE state. The magnetic structure of the different phases was studied by elastic neutron scattering on single crystals using the triple-axis spectrometers HB1 and HB1A at the High-Flux Isotope Reactor, Oak Ridge National Laboratory. Different scattering geometries were used in order to probe the competing magnetic phases. Horizontal collimations of 48'-60'-60'-120' were employed and the final neutron energy was fixed at 13.5 meV.

\begin{figure}
\begin{center}
\includegraphics[angle=0,width=3in]{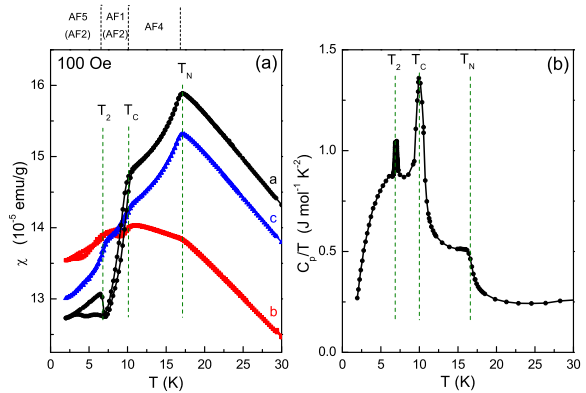}
\caption{(Color online) (a) Magnetic susceptibility and (b) heat capacity of Mn$_{0.85}$Co$_{0.15}$WO$_4$. Three phase transitions are indicated by their critical temperatures T$_N$, T$_C$, and T$_2$, respectively, upon decreasing temperature. The major phases are labeled at the top of (a). The AF2 phase (in brackets) coexists with AF1 and AF5.}
\end{center}
\end{figure}

\section{Results and Discussion}
\subsection{Magnetic susceptibility and heat capacity}
Fig. 1a shows the low-field magnetic susceptibility measured along the three crystallographic orientations. The onset of magnetic order at T$_N\approx$17 K is marked by a sudden decrease of the a- and c-axis susceptibilities ($\chi_a$, $\chi_c$) and a distinct slope change of $\chi_b$. The value of T$_N$ is in good agreement with the phase boundary of the AF4 phase derived from powder data.\cite{song:09} At T$_C\approx$ 10 K there is another anomaly most pronounced as a sharp drop of $\chi_a$, but also well resolved in $\chi_b$ and $\chi_c$. The transition temperature coincides with the AF4 $\rightarrow$ (AF2'+AF4) phase boundary of Ref. \cite{song:09}. At lower temperature, however, the susceptibility data reveal a third phase transition near T$_2$ = 7 K, not reported before. Both low temperature phases exhibit a pronounced thermal hysteresis (Fig. 1a). The three phase transitions are confirmed by heat capacity measurements (Fig. 1b). C$_p$(T) exhibits distinct anomalies at the critical temperatures where the anomalies of $\chi$ have been detected (dashed lines in Fig. 1).

\subsection{Dielectric constant and ferroelectric polarization}
In most multiferroic compounds the dielectric constant exhibits well defined anomalies at the magnetic (and ferroelectric) phase transitions. The b-axis dielectric constant of Mn$_{0.85}$Co$_{0.15}$WO$_4$ is shown in Fig. 2a. The onset of magnetic order is indicated by a sudden change of slope at T$_N$ followed by a continuous decrease of $\varepsilon_b$ in the AF4 phase. At T$_c$, $\varepsilon_b$ shows a very sharp and pronounced peak indicative of a possible ferroelectric transition with a spontaneous b-axis polarization and another, smaller peak at T$_2$. The dielectric loss measured along the b-axis (Fig. 2b) exhibits two sharp peaks at the same temperatures as is expected near FE phase transitions. The critical temperatures of the two peaks coincide with the anomalies observed in the magnetization (Fig. 1a) and heat capacity (Fig. 1b). It is interesting that the second peak of $\varepsilon_b$ and the loss factor seem to indicate a second FE transition at T$_2$. The neutron scattering data, discussed below, confirm the formation of another magnetic structure at T$_2$.

In contrast to the b-axis dielectric data, measurements of the a-axis dielectric constant have not shown any similar sharp peak that might hint for the existence of an a-axis FE polarization as suggested earlier.\cite{song:09} Instead, at the critical temperatures T$_c$ and T$_2$ a distinct change of slope of $\varepsilon_a$ can be observed. The dielectric loss along the a-axis (Fig. 2b) does not display any peak or other anomaly near T$_c$ or T$_2$. The absence of sharp peaks of the a-axis dielectric constant and loss factor makes the existence of a significant a-axis component of the polarization exists in the FE state very unlikely.

\begin{figure}
\includegraphics[angle=0,width=3in]{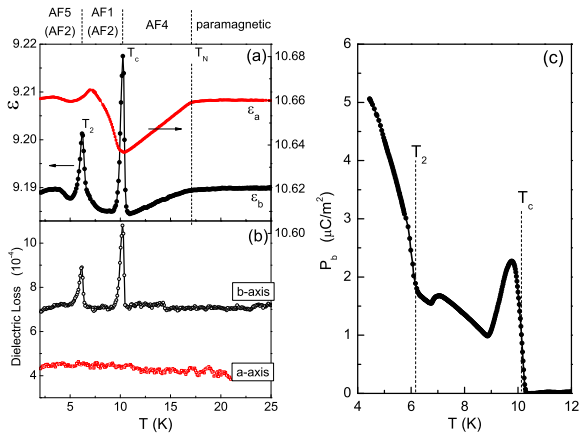}
\caption{(Color online) (a) Dielectric constant, (b) dielectric loss, and (c) ferroelectric polarization of Mn$_{0.85}$Co$_{0.15}$WO$_4$ measured along the b-axis. The magnetic phases are labeled as in Fig. 1.}
\end{figure}

To prove the existence of ferroelectricity in Mn$_{0.85}$Co$_{0.15}$WO$_4$ and to identify the orientation of the polarization the pyroelectric current was measured along the a- and b-axes. The a-axis measurement could not detect any pyroelectric current within the resolution of the experiment ruling out a significant a-axis component of the electric polarization. The b-axis measurement, however, does reveal a pyroelectric current with a narrow peak at T$_c$ consistent with a sharp rise of the electric polarization. Fig. 2c displays the resulting b-axis polarization. P$_b$(T) rises to a maximum below T$_c$, decreases slightly toward lower T, and increases again below 8.5 K upon further decreasing temperature. A sharp rise of P$_b$ below T$_2$ is consistent with the sharp peak of $\varepsilon_b$ and the loss factor at this temperature and it supports the conjecture of a second FE transition. This behavior indicates a possible competition and coexistence of different magnetic phases as for example in Mn$_{1-x}$Fe$_x$WO$_4$.\cite{ye:08,chaudhury:09b} The coexistence of two magnetic phases (AF2' and AF4) below 10 K was also suggested from powder neutron measurements.\cite{song:09} The magnitude of P$_b$$\approx$5 $\mu$C/m$^2$ of Mn$_{0.85}$Co$_{0.15}$WO$_4$ is significantly lower than typical values for MnWO$_4$ which could be attributed to the high degree of cobalt substitution and the coexistence of the multiferroic state with competing paraelectric magnetic phases.

\subsection{Neutron scattering results}
Magnetization and dielectric measurements do not reveal the details of the magnetic orders of the various phases observed. Neutron scattering experiments on single crystals of Mn$_{0.85}$Co$_{0.15}$WO$_4$ have therefore been conducted with emphasis on the modulation (IC or CM) of the magnetic order parameter, possible coexistence of phases, and the intensity of the reflections. Figs. 3a,b show the characteristic pattern obtained for two basic scattering geometries. The color intensity map (Fig. 3a) for $\overrightarrow{q}=(H,0.5,L)$ with $L=-2H$ shows different CM and IC peaks labeled "AF1", "AF2", and "AF5", respectively. A more precise determination of the magnetic modulation vectors associated with the IC diffraction peaks requires to determine the peak position for H and L independently, as shown in Figs. 3c and 3d at temperatures of 9.8 K (AF2) and 6.0 K (AF5). This leads to the following values of $\overrightarrow{q}$ for the IC AF2 and AF5 phases: $\overrightarrow{q}_{AF2}=(0.214\pm0.002,0.5,-0.457\pm0.003)$ and $\overrightarrow{q}_{AF5}=(0.228\pm0.002,0.5,-0.480\pm0.003)$. $\overrightarrow{q}_{AF2}$ is the same as in the undoped MnWO$_4$ that shows electric polarization and $\overrightarrow{q}_{AF5}$ is distinctly different from $\overrightarrow{q}_{AF2}$.

The neutron scattering data show a more complex picture of the magnetic orders below T$_N$. At high temperature, between T$_N$ and T$_c$, the AF4 magnetic order with modulation vector $\overrightarrow{q}_{AF4}=(1/2,0,0)$ is the only magnetic phase (Fig. 3b). Below T$_c$ the intensity of the AF4 phase decreases significantly but some small intensity can still be observed to the lowest temperatures, consistent with Ref. \cite{song:09}. However, at T$_c$ the strong reflections of the AF1 phase with $\overrightarrow{q}_{AF1}=(\pm1/4,1/2,1/2)$ appear and the AF1 phase is the major phase between T$_c$ and T$_2$ (Fig. 3a), in contrast to the recent results.\cite{song:09} The intensity of the AF1 reflections are largely diminished below T$_2$ where an IC phase with wave vector $\overrightarrow{q}_{AF5}=(0.228,0.5,-0.480)$ becomes visible. While the AF5 phase is the major phase below T$_2$ the neutron spectra also show intensities of the AF4, the AF1, and the AF2 phases. The AF2 phase with $\overrightarrow{q}_{AF2}=(0.214,0.5,-0.457)$ does coexist with the other magnetic phases below T$_c$ but its reflections are relatively weak and it seems to be a minor phase.

\begin{figure}
\includegraphics[angle=0,width=3in]{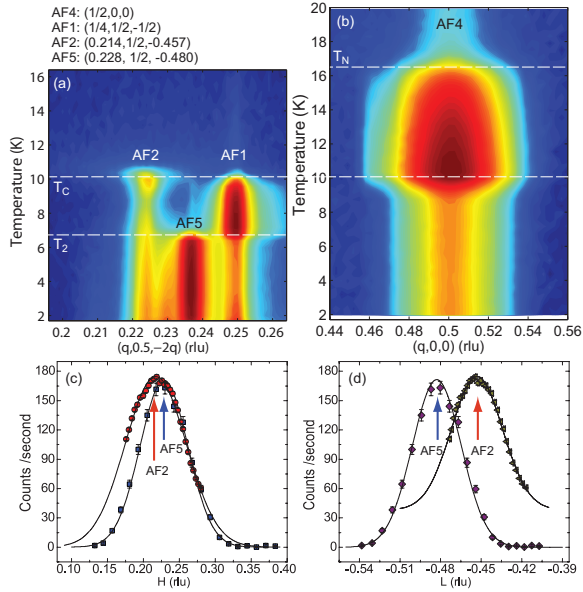}
\caption{(Color online) Neutron scattering intensities vs temperature probing the different magnetic structures of Mn$_{0.85}$Co$_{0.15}$WO$_4$. (a) Crystal alignment in the plane defined by (1,0,-2) and (0,1,0). (b) Crystal alignment in the plane defined by (1,0,0) and (0,1,0). (c) and (d) Refinement of the magnetic modulation vectors of phases AF2 (T=9.8 K) and AF5 (T=6 K).}
\end{figure}

\begin{figure}
\includegraphics[angle=0,width=3in]{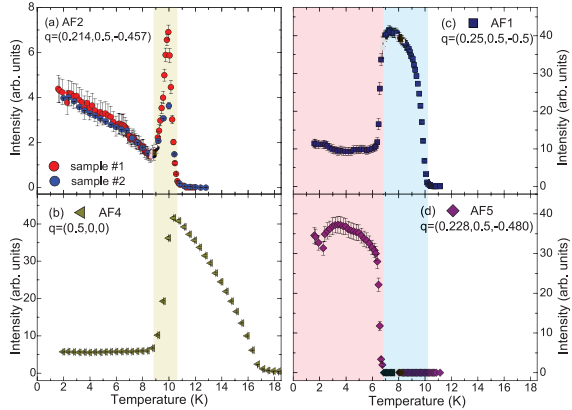}
\caption{(Color online) Integrated intensity of magnetic peaks of the (a) ICM AF2 phase, (b) CM AF4 phase, (c) CM AF1 phase, and (d) ICM AF5 phase. Only the CM AF4 phase persists above 10 K, while various competing magnetic orders coexist at lower temperature. Two different single crystals are measured to ensure the complex transitions are intrinsic to the nominal Co concentration.}
\end{figure}

Following the development of the neutron scattering reflections with the highest intensities the phase sequence of the major magnetic orders upon decreasing temperature is AF4 $\rightarrow$ AF1 $\rightarrow$ AF5. While the IC AF5 phase could be identical to the AF2' phase noted in powder neutron experiments\cite{song:09} the main difference to the previous work is the unambiguous detection of the AF1 phase in an intermediate temperature range. In addition, it is obvious that the ferroelectric AF2 phase coexists with AF5 between T$_c$ and the lowest temperature. This raises the question whether the observed b-axis polarization (Fig. 2b) can solely be ascribed to the AF2 phase.

The coexistence of different magnetic structures below T$_c$ is clearly shown in Fig 3a and the integrated intensities displayed in Fig. 4. The sudden change of intensities of the major scattering peaks with temperature defines the three magnetic phase transitions in good agreement with the data shown in Figs. 1 and 2. The intensity of the AF2 reflection from Fig. 4a has a very similar temperature dependence as the FE polarization P$_b$. The peak below the ferroelectric transition temperature T$_c$, the minimum at 8.5 K, and the smooth increase to lower temperature mimic P$_b$(T) shown in Fig. 2c. This suggests that the observed b-axis polarization just below T$_c$ but above T$_2$ arises from the AF2 phase.

The peculiar T-dependence of the AF2 scattering intensity indicates a strong competition of this phase with the commensurate AF4 and AF1 phases. Above T$_c$ only the AF4 phase is stable (Fig. 4b). At T$_c$ the AF4 phase weakens and the AF1 phase arises (Fig. 4c). Simultaneously, the AF2 phase shows up with an intensity peak right below T$_c$ (Fig. 4a). When the AF1 commensurate phase grows stronger at slightly lower temperature the AF2 intensity is quickly reduced resulting in the sharp intensity peak at T$_c$. This is a result of the competition between AF1 and AF2 as is well known for the undoped compound MnWO$_4$ and from a simple model calculation.\cite{chaudhury:08} At the lower temperature T$_2$ the AF5 phase grows at the expense of AF1. However, it is not clear whether the IC AF5 phase contributes to the observed FE polarization. The rapid increase of P$_b$(T) near T$_2$ where the AF5 scattering intensity starts to increase does indicate a possible contribution of the AF5 phase to P$_b$. Adding the intensities if both, AF2 and AF5, results in qualitatively similar T-dependence as is shown by the FE polarization (Fig. 2b). Therefore, it seems conceivable that two magnetic orders, AF2 and AF5, contribute to the FE polarization along the b-axis. The second sharp peak of $\varepsilon_b$ near 6 K (Fig. 2a) and the similar peak of the dielectric loss lends further support to this conjecture.

It appears interesting that the sequence of the major phases in Mn$_{0.85}$Co$_{0.15}$WO$_4$ with increasing T is IC(AF5)$\rightarrow$CM(AF1)$\rightarrow$CM(AF4). The transition from AF5 to AF1 can be considered as an inverse lock-in transition. The theory of simple frustrated magnetic systems usually explains the lock-in transition from a commensurate ground state to an incommensurate phase at higher temperatures. The reversal of the lock-in property of the magnetic order in Mn$_{0.85}$Co$_{0.15}$WO$_4$ is a result of the complexity of frustrated magnetic exchange interactions and anisotropy that leads to the unconventional phase sequence. This suggestion is qualitatively consistent with the conclusion from inelastic neutron scattering experiments on MnWO$_4$ that the magnetic coupling of up to the ninth neighbors is needed to explain the spin wave dispersion spectrum.\cite{ehrenberg:99} A similar IC$\rightarrow$CM sequence has previously been reported in RMn$_2$O$_5$ multiferroics, a materials system that also shows a very complex magnetic phase diagram with competing and frustrated magnetic exchange interactions.\cite{kobayashi:04b}

\section{Summary and Conclusions}
The successful growth of large single crystals of the solid solution Mn$_{0.85}$Co$_{0.15}$WO$_4$ and their investigation by means of magnetic, heat capacity, dielectric, polarization, and neutron scattering experiments has revealed a very complex magnetic and multiferroic phase diagram. Three phase transitions have been identified upon decreasing temperature. The major phase sequence, AF4 $\rightarrow$ AF1 $\rightarrow$ AF5, includes two commensurate orders and a low-temperature incommensurate phase. A recent suggestion that this phase may be associated with a large a-axis ferroelectric polarization could not be confirmed. Several phases coexist in different temperature ranges and, below 10 K, the characteristic reflections of the ferroelectric helical AF2 phase have been observed with relatively low intensity. The ferroelectric polarization is finite in the same temperature range and its amplitude closely tracks the T-dependence of the AF2 scattering intensity. A sudden increase of the b-axis polarization at low temperature is ascribed to an additional contribution from the incommensurate AF5 magnetic structure, stable below 7 K. At the lowest temperature four phases seem to coexist with the largest scattering intensity arising from the AF5 phase. The specific nature of this phase is currently being investigated. The phase sequence observed in Mn$_{0.85}$Co$_{0.15}$WO$_4$ suggests that the phase diagram of the solid solution Mn$_{1-x}$Co$_{x}$WO$_4$ shows considerably more complexity than previously assumed. The x-dependence of the magnetic phases and the complete magnetic and multiferroic phase diagram of Mn$_{1-x}$Co$_{x}$WO$_4$ is a matter of forthcoming studies.

\acknowledgments
This work is supported in part by the T.L.L. Temple Foundation, the J.J. and R. Moores Endowment, the State of Texas through TCSUH, the USAF Office of Scientific Research, at LBNL through the US Department of Energy, and by the Division of Scientific User Facilities of the Office of Basic Energy Sciences, US Department of Energy.



\end{document}